\documentclass[usenatbib]{mnras}

\usepackage[T1]{fontenc}
\usepackage{ae,aecompl}

\usepackage{graphicx}	% Including figure files
\usepackage{amsmath}	% Advanced maths commands
\usepackage{amssymb}	% Extra maths symbols

\title[Variable Stars in Quintuplet]{Variable Stars in the Quintuplet stellar cluster with the VVV Survey}

\author[C. Navarro Molina et al.]{
Claudio Navarro Molina,$^{1,2}$\thanks{E-mail: claudio.navarro@postgrado.uv.cl}
J. Borissova,$^{2,1}$
M. Catelan,$^{3,1}$
J. Alonso-Garc\'{\i}a,$^{4,1}$
\newauthor E. Kerins,$^{5}$
R. Kurtev,$^{2,1}$
P. W. Lucas,$^{6}$
N. Medina,$^{1,2}$
D. Minniti,$^{7,8,1}$
and I. D\'ek\'any,$^{1,3}$
\\
$^{1}$ Millennium Institute of Astrophysics, Vicu\~na Mackenna 4860, 782-0436 Macul, Santiago, Chile\\
$^{2}$ Instituto de Astrof\'{\i}sica, Universidad de Valpara\'{\i}so, Gran Breta\~na 1111, Casilla 5030, Valpara\'{\i}so, Chile\\
$^{3}$ Instituto de Astrof\'{\i}sica, Facultad de F\'{\i}sica, Pontificia Universidad Cat\'olica de Chile, Casilla 306, Santiago 22, Chile\\
$^{4}$ Unidad de Astronom\'{\i}a, Facultad de Ciencias B\'asicas, Universidad de Antofagasta, Avenida U. de Antofagasta 02800, Antofagasta, Chile\\
$^{5}$ Jodrell Bank Centre for Astrophysics, University of Manchester, Manchester M13 9PL, UK\\
$^{6}$ Centre for Astrophysics Research, Science and Technology Research Institute, University of Hertfordshire, Hatfield, AL10 9AB, UK\\
$^{7}$ Departamento de Ciencias F\'{\i}sicas, Universidad Andr\'es Bello, Rep\'ublica 220, Santiago, Chile\\
$^{8}$ Vatican Observatory, V00120 Vatican City State, Italy\\
}

\date{Accepted XXX. Received YYY; in original form ZZZ}

\pubyear{2016}

\begin{document}
\label{firstpage}
\pagerange{\pageref{firstpage}--\pageref{lastpage}}
\maketitle

\begin{abstract}
The Quintuplet cluster is one of the most massive star clusters in the Milky Way, situated very close to the Galactic center. We present a new search for variable stars in the vicinity of the cluster, using the five-year database of the Vista Variables in the Via Lactea (VVV) ESO Public Survey in the near-infrared. A total of 7586 objects were identified in the zone around $2'$ from the cluster center, using 55 $K_S$-band epochs. Thirty-three stars show $K_S$-band variability, 24 of them being previously undiscovered. Most of the variable stars found are slow/semiregular variables, long-period variables of the Mira type, and OH/IR stars. In addition, a good number of our candidates show variations in a rather short timescale. We also propose four Young Stellar Object (YSO) candidates, which could be cluster members.
\end{abstract}


\section{Introduction}

The Galactic Center (GC) is an exceptional laboratory to test the formation and evolution of stars under extreme conditions. Three massive and young star clusters have been found in this region: Arches, Quintuplet and the central cluster surrounding Sagittarius A*, a supermassive black hole. Each is supposed to contain about $10^4$ stars. Among these, Quintuplet has been known to be rich in massive stars, such as Wolf-Rayet and OB supergiants \citep{fig99}.

Even though this is a well-studied cluster, not many variability searches have been carried out, and the few available are mostly restricted to bright stars. This could be a consequence of the observational challenges this field poses. The first known variable was the \lq Pistol star\rq \, \citep{fig98}, a luminous blue variable (LBV), and one of the brightest stars in the Milky Way; a second LBV was discovered by \citet{geb99}. \citet{gla99} found several variable candidates brighter than $K_S=11$, including some large-amplitude asymptotic giant branch (AGB) variables. Later on, \citet[hereafter MKN09]{mat09} performed a near-infrared survey of Miras in the GC, which covered the Quintuplet area.

It is clear that a better understanding of the stellar population of this cluster can provide valuable information about the life cycle of stars in the central regions of our Galaxy. The irruption of near-infrared surveys, like the Vista Variables in the Via Lactea ESO Public Survey \citep[VVV;][]{min10, sai12}, can help us to achieve that goal, due to its multi-epoch and multi-band nature. For example, \citet{dek15} recently discovered a young thin stellar disk population across the Galactic bulge, using data from this survey.

In this paper we analyze $K_S$-band data from the VVV survey around the Quintuplet star cluster, in order to find variable sources that allow us to get more insights about the evolution of stars in the cluster.

\subsection{Quintuplet}

The Quintuplet cluster ($\alpha$: $17^h46^m15^s$, $\delta$: $-28^{\circ}49'41''$; J2000), named after the prominent presence of five bright stars \citep{nag90, oku90}, is located at a projected distance of about 30~pc from Sgr A*. With an age of $3.0\pm0.5$~Myr \citep{lie12, lie14}, and a tidal radius of $\approx1$~pc \citep{por02}, the central part of the cluster presents a flat present-day mass function with a slope of $-1.68$ \citep{hus12}, which may be caused by the fast dynamical evolution of the cluster within the strong Galactic tidal field. A similar result has been found for the inner region of Arches \citep{hab13}.

Given that massive clusters in the GC are supposed to dissolve within a few tens of Myr \citep{por02}, we expect that these clusters significantly contribute to the isolated massive stars population \citep{hab14}. Furthermore, \citet{hus12} determined cluster membership through proper motion analysis. And finally, \citet{sto14} studied the orbital motion of the Quintuplet, and suggested that this cluster and Arches may have a common origin, a situation that we intend to address in a subsequent paper.

%--------------------------------------------------------------------------------------------------------------

\section{The Data Set}

To investigate the cluster area we used images from the VVV ESO Public Survey \citep{min10,sai12}. This survey has been carried out with the 4.1~m VISTA telescope at Cerro Paranal, Chile, equipped with the VIRCAM camera, an array of 16 chips, each with a resolution of 2048x2048 pixels. VVV covers a total area of $562$~deg$^2$, corresponding to the bulge of our Galaxy and a section of the mid-plane. Every single exposure produced is called a {\it pawprint}. In order to fill the gaps between the detectors, a pattern of 6 images with different offsets is required. The combination of all offsets is called a {\it tile}. Each tile has a field of $1.6$~deg$^2$, meaning that the VVV requires 348 tiles to cover the survey area. 

Instead of using the full tile images, we opted to work with the individual stacked pawprints, in order to avoid photometry problems due to a highly variable point-spread function (PSF), and mosaicing issues \citep{alo15}. For this work we used tile b333, which corresponds to the GC; more specifically, chip 15, which contains the Quintuplet cluster, covering a region of $11.5' \times 11.5'$. From the available epochs, we selected only those with obstatus=`Completed' and QC (quality control) grade=`A', meaning only the best quality images were used. Additionally, epochs with a seeing higher than 1 arcsec ($\sim 3$ pixels) were discarded. Since the offset pattern ensures that every region is observed at least twice, we decided to use both sets of images (which we call set A and set B) to compare their photometry (see Section~\ref{method} for further details). The already mentioned selection criteria reduced the number of epochs to 55 and 56 from sets A and B, respectively, which corresponds to a time interval between August 2010 and May 2014. Data were reduced, astrometrized and calibrated using the Cambridge Astronomical Survey Unit (CASU) pipeline, version 1.3 \citep{irw04}.

%----------------------------------------------------------------------

\section{Photometry}

As a first approach, we directly analyzed the photometric catalogs from CASU. However, we realized that for our purposes, aperture photometry was not reliable, due to the strong crowding, even in the external zones of the cluster. Thus, we used DoPHOT \citep{sch93, alo12} to perform PSF photometry of the $K_S$ images. Input parameters, like seeing, average sky level, gain and readout noise, were extracted directly from each image header. Details of the observational conditions can be found on the CASU website\footnote{http://casu.ast.ca.ac.uk/vistasp/}. From the catalogs obtained, only objects with a {\it chi} value lower than 3, and a DoPHOT flag of 1 or 7 were kept\footnote{A DoPHOT flag of 1 corresponds to the stars with the best photometry; flag 7 means that stars are faint, but photometry is still reliable.}. This selection allows us to keep the best quality objects for each epoch. Additional $J$ and $H$-band single epochs were reduced in the same way as $K_S$-band images, and allowed us to build a color-magnitude diagram for the cluster region.

\subsection{Photometric calibration}

\begin{figure}
\includegraphics[width=\columnwidth]{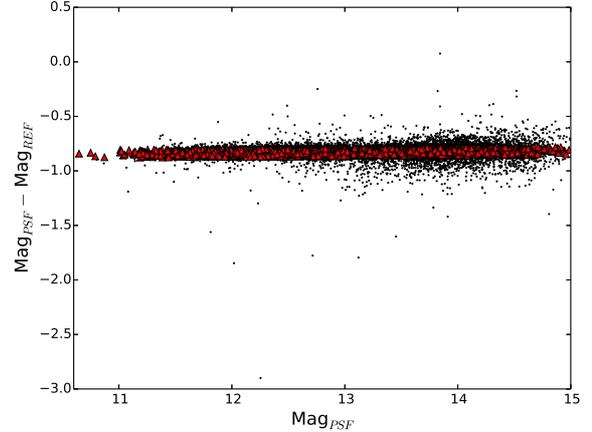}
\caption{Linear fit performed to obtain the transformation equation for a single epoch. Mag$_{PSF}$ are the instrumental magnitudes obtained directly from DoPHOT, while Mag$_{REF}$ are the magnitudes of these objects in the CASU catalogs. Small dots correspond to the full sample, while red triangles are the remaining objects after the 5-sigma clipping.\label{fig:linfit}}
\end{figure}

DoPHOT catalogs were calibrated by searching for a list of non variable, isolated stars within the chip field in the aperture photometry catalogs provided by CASU. Only stars with $11 < K_S < 15$ were selected. With a total of 1280 standard stars, we performed a linear fit and a 5-sigma clipping in order to get the transformation equations for each epoch. Figure~\ref{fig:linfit} shows the linear fit for a single epoch. A similar procedure was carried out to calibrate the $J$ and $H$ images. Further details such as the flux-to-magnitude conversion, calibration of zero points and color terms, can be found in the ESO website\footnote{https://www.eso.org/sci/observing/phase3/data\textunderscore releases/vvv\textunderscore dr1.html}.

\subsection{Completeness Test}
\label{compl}

\begin{figure}
\includegraphics[width=\columnwidth]{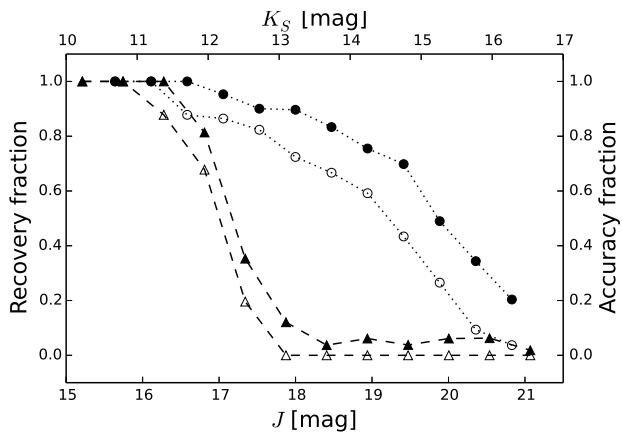}
\caption{Recovery (full symbols) and accuracy (open symbols) fraction of artificial stars added to the cluster image (see text for details). Circles correspond to the $K_S$-band, while triangles are the results for the $J$-band. Completeness falls much faster for J-band due to the strong reddening.\label{fig:recov}}
\end{figure}

In order to derive the photometric completeness, we added artificial stars using the IRAF task {\it mkobjects}. A total of 600 stars were added to the first epoch image, with $K_S$ magnitudes between 11.5 and 16.5, with a uniform luminosity function, meaning that we created 50 stars per each bin of 0.5 magnitudes. Then we checked how many stars we recovered. In addition, we also computed the fraction of stars whose measured magnitude was similar, within a three-sigma level, to the original value. This is what we called the {\it accuracy} fraction. This procedure was repeated 5 times with a different spatial distribution, and the same test was carried out for the $J$-band. Results are shown in Figure~\ref{fig:recov}. Here we can conclude that the photometry is complete and accurate, at the 80\% level, for magnitudes up to $K_s \approx 13$ and $J \approx 16.5$.

\subsection{DIA}

Differential Image Analysis (DIA; Wozniak 2008 and references therein) is a useful technique for variability detection in crowded fields. In this method, a reference image is selected whether as a single epoch from a data set, or as a combination of several images. All other images are transformed to the coordinate system of the also called \lq template\rq. Then the reference image is convolved by a kernel function, and finally every single epoch of the data set is subtracted from this convolved image \citep[e.g.,][]{ala98,ala00}. Usually the template turns out to be the best seeing image available. However, \citet{huc14} found that for the particular case of the VVV survey, a better sampled image (i.e. with a poorer seeing) is more suitable as a reference.

DIA was applied to all the epochs available for the same region used with DoPHOT, this is, sets A and B of tile b333, chip 15. Light curves obtained were then analyzed using the methods that will be described in the next subsection.

%---------------------------------------------------------------

\section{Results}

\subsection{Candidate Selection}
\label{method}

From the catalog corresponding to chip 15, we selected stars in a radius of $2'$ around the cluster center. This means a total of 7596 objects. We decided to go beyond the tidal radius to analyze the cluster surroundings and compare the stars found with the assumed cluster population. In order to search for variability, we implemented two selection criteria. The first one looked for strong variations in the light curves:

\begin{displaymath}
I = \sqrt{\frac{1}{N_{epochs}}\sum{\left( \frac{\Delta M_i}{\sigma(M_i)} \right) ^2}}
\end{displaymath} 

$\Delta M_i$ is the difference between the magnitude on a single epoch, and the expected value, calculated from a linear fit performed to the light curve. $\sigma(M_i)$ is simply the photometric error, as returned by DoPHOT. This criterion was chosen, instead of the other common variability indexes found in the literature, since we were originally searching for YSOs with mostly irregular light curves. Selecting objects with $I>5$ allows us to remove variations that are within the photometric errors. This script was performed once for each set available. After keeping light curves with more than 25 epochs, the list of candidates present in both image sets was reduced to 37.

\begin{figure}
\includegraphics[width=\columnwidth]{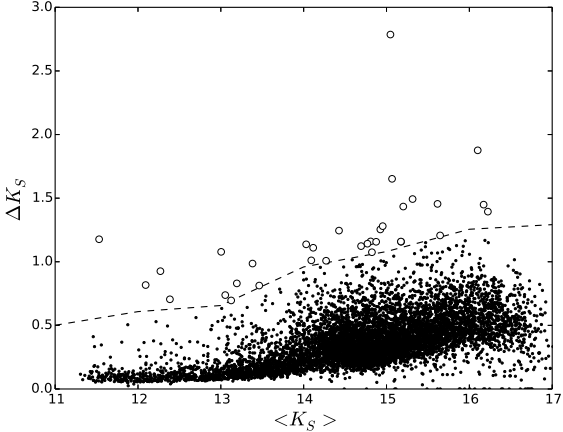}
\caption{Amplitude versus mean magnitude for all the targets found with DoPHOT. The dashed line corresponds to 4-sigma above the mean amplitude (see text for details). Selected candidates are represented as open circles.\label{fig:carlos}}
\end{figure}

For the second method \citep{con15} we calculated the mean amplitude (defined as the difference between the minimum and maximum magnitude) per magnitude bin for our sample. Then we selected targets that were 4-sigma above the mean (Figure~\ref{fig:carlos}). With this method we found 33 candidates; however, only 15 were found in common between the sets A and B. Of these, 10 candidates were already found using the first criterion. 

Since the time difference between two consecutive pawprint observations is less than one minute, meaning a timescale much shorter than the typical for brightness variations such as we are interested in, we could safely compare the two light curves available per candidate, in order to account for spurious detections due to image artifacts, or problems with the photometry, under the assumption that both curves should be similar, within the photometric uncertainties. After a visual inspection of the light curves, 11 candidates were discarded, mainly due to poor correlation between the pawprints or blending. With this last selection, our final list of candidates found with DoPHOT contained 31 stars.

\begin{figure}
\includegraphics[width=\columnwidth]{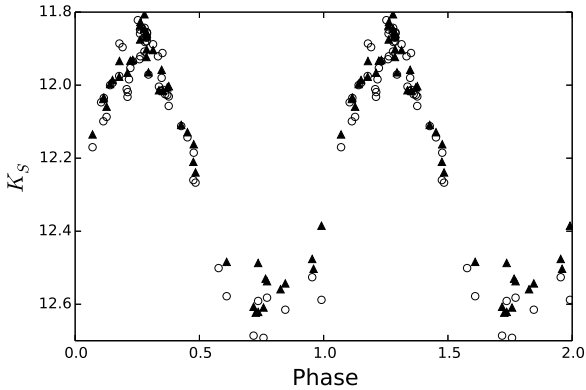}
\caption{Comparison of light curves obtained with the two techniques used in this paper, DIA (open circles) and DoPHOT (filled triangles), for the variable candidate VC19. A period of 110.9 days was used to construct the phased diagram. Both curves are nearly similar, within typical photometric uncertainties.\label{fig:dia-doph}}
\end{figure}

\begin{figure*}
\includegraphics[width=\textwidth]{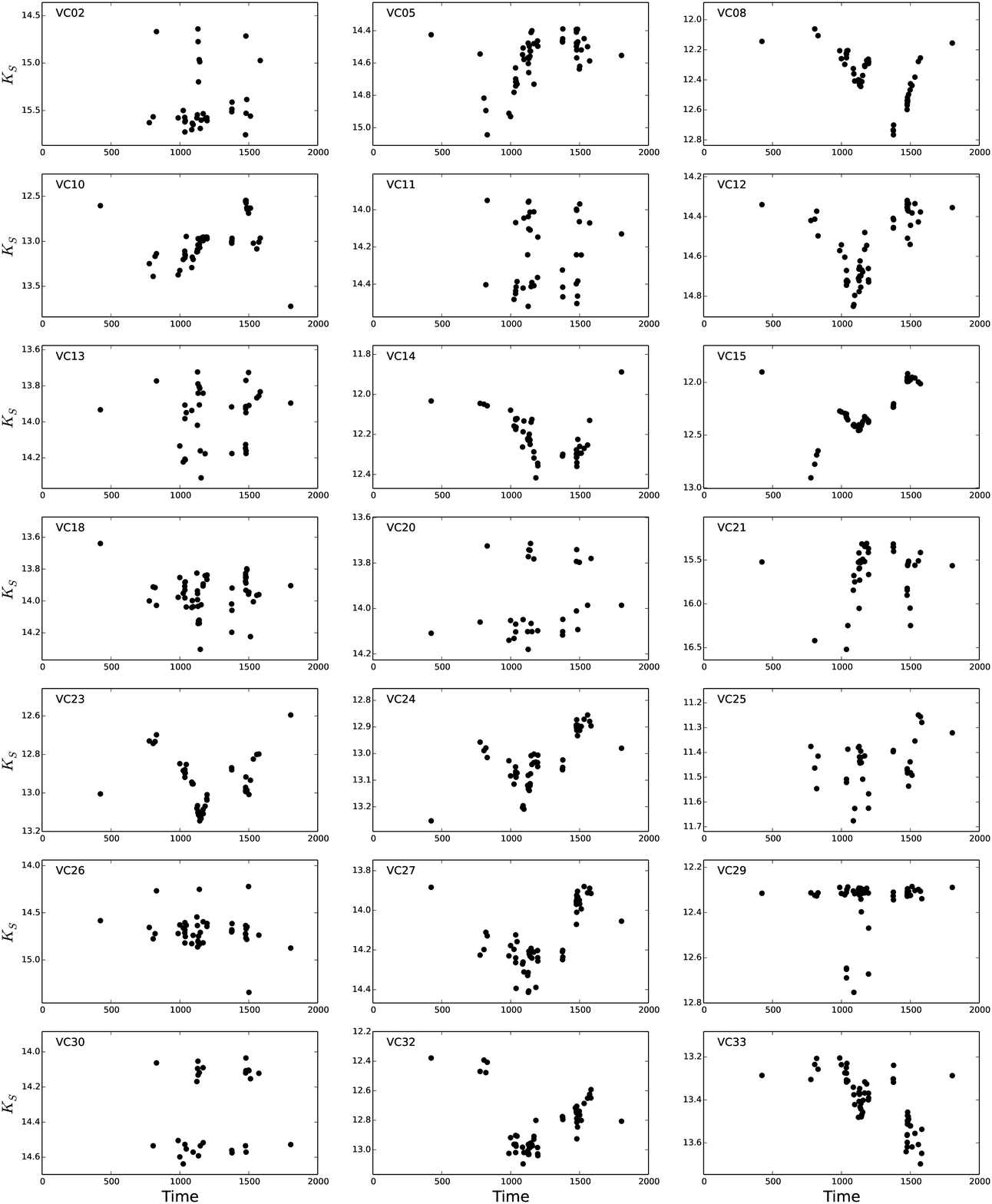}
\caption{Light curves for non periodic variable candidates. Time is $MJD-55000$ days.\label{fig:nonper}}
\end{figure*}

A similar procedure was carried out with the light curves obtained with DIA. We found 5 objects, however, three of them were already present in the DoPHOT catalog (VC07, VC17 and VC19). The smaller number of candidates obtained was due to the fact that this version of DIA did not perform optimally in fields with very high levels of crowding, as in the case here. Figure~\ref{fig:dia-doph} shows a comparison between the light curves obtained with DoPHOT (triangles) and DIA (open circles) for VC19. A good agreement is generally observed between both techniques \citep[see also][]{cat13}, although for simplicity, we decided to use their DoPHOT light curves for the remaining analysis. With all these selection criteria applied, the final list of candidates was 33. Table~\ref{tbl-1} contains the main features of these candidates, including their name, coordinates, mean $K_S$ magnitude, and $\Delta K_S$, plus a single epoch $J$ and $H$ magnitude, and GLIMPSE $[3.6]$ and $[4.5]$-band magnitudes, if applicable. Finding charts for all candidates can be found in Appendix~1. Variables 1 through 31 were found using DoPHOT, while 32 and 33 were obtained with DIA. Figure~\ref{fig:nonper} contains the light curves for non-periodic candidates.

\begin{figure*}
\includegraphics[width=\textwidth]{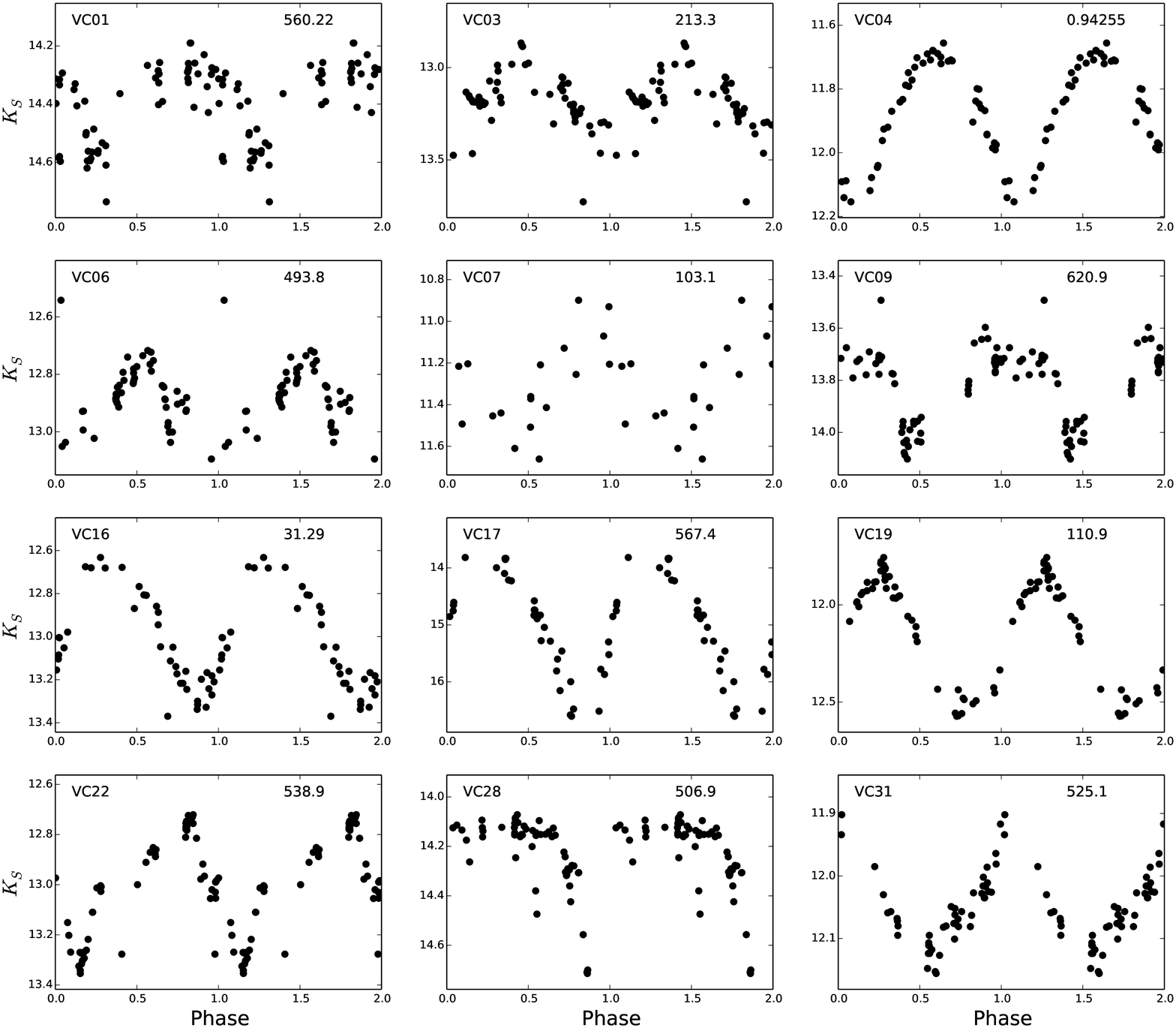}
\caption{Phased light curves for periodic variable candidates\label{fig:per}}
\end{figure*}

Some of the variable candidates showed signs of periodicity. In order to estimate the time interval of their variations, we used a generalized Lomb-Scargle periodogram \citep{sca82}, imposing a false-alarm probability cutoff at 1\% to help ensure the reliability of the period found. In total, twelve variables had periodicity. Last column of Table~\ref{tbl-1} displays the periods found, while Figure~\ref{fig:per} shows the corresponding phased light curves.

%\begin{deluxetable}{rcccccccc}
%\tablecolumns{6}
%\tablewidth{0pt}
\begin{table*}
\centering
\caption{Variable Candidates found with DoPHOT and DIA}
\label{tbl-1}
\begin{tabular}{rccccccccc}
\hline
ID & $RA_{J2000}$ [deg] & $DEC_{J2000}$ [deg] & $J$ & $H$ & $\langle K_S \rangle$ &  $[3.6]$ &  $[4.5]$ &  $\Delta K_S$ &  $P$ [days]\\

\hline
VC01 & 266.5665557 & -28.8513395 & & 17.042 & 14.403 &  &  & 0.546 & 560.2 \\
VC02 & 266.5531019 & -28.8423316 & & 17.382 & 15.413 &  &  & 1.116 &  \\
VC03 & 266.5547296 & -28.8365616 & 17.701 & 15.282 & 13.182 &  &  & 0.86 & 213.3 \\
VC04 & 266.5488075 & -28.8333713 & 15.994 & 13.522 & 11.861 & 11.127 & 10.361 & 0.498 & 0.94255 \\
VC05 & 266.5385634 & -28.8224738 & & 17.469 & 14.572 &  &  & 0.655 &  \\
VC06 & 266.5866369 & -28.8473154 & & 17.600 & 12.872 & 10.816 & 9.636 & 0.553 & 493.8 \\
VC07 & 266.5391027 & -28.8216129 & 17.510 & 13.927 & 11.576 & 9.482 & 9.171 & 1.193 & 103.1 \\
VC08 & 266.5680057 & -28.8331950 & & 15.855 & 12.385 &  &  & 0.705 &  \\
VC09 & 266.5842469 & -28.8401378 & & 17.771 & 13.819 & 11.515 & 10.827 & 0.609 &  \\
VC10 & 266.5835128 & -28.8368835 & & 15.127 & 12.980 & 11.095 & 10.542 & 1.176 &  \\
VC11 & 266.5468613 & -28.8166420 & & 16.024 & 14.241 &  &  & 0.569 &  \\
VC12 & 266.5720312 & -28.8235543 & & 17.220 & 14.548 &  &  & 0.532 &  \\
VC13 & 266.5392703 & -28.8443353 & 19.032 & 15.755 & 13.973 &  &  & 0.588 &  \\
VC14 & 266.5701077 & -28.8189804 & 19.244 & 14.507 & 12.218 &  &  & 0.529 &  \\
VC15 & 266.5444024 & -28.8038501 & & 16.084 & 12.280 & 10.099 & 8.780 & 1.004 &  \\
VC16 & 266.5565088 & -28.8097095 & 19.262 & 15.124 & 13.050 & 10.078 & 9.943 & 0.738 & 31.29 \\
VC17 & 266.5639319 & -28.8133190 & & & 15.047 & 7.397 & 6.045 & 2.786 & 567.4 \\
VC18 & 266.5889574 & -28.8250212 & & 16.643 & 13.955 & 12.137 & 11.657 & 0.665 &  \\
VC19 & 266.5765322 & -28.8156985 & & 15.604 & 12.092 & 10.772 & 9.656 & 0.817 & 110.9 \\
VC20 & 266.5468932 & -28.7998864 & 18.497 & 15.654 & 13.973 & 12.030 & 11.600 & 0.466 &  \\
VC21 & 266.5883325 & -28.8212281 & & & 15.647 &  &  & 1.207 &  \\
VC22 & 266.5945312 & -28.8231272 & & 16.421 & 13.011 & 11.275 & 10.512 & 0.633 & 538.9 \\
VC23 & 266.5617898 & -28.8049052 & & 16.928 & 12.954 & 11.724 & 10.554 & 0.55 &  \\
VC24 & 266.5919313 & -28.8198884 & & 16.641 & 13.02 & 11.594 & 11.021 & 0.398 &  \\
VC25 & 266.5895737 & -28.8114513 & 19.066 & 13.930 & 11.446 & 9.664 & 9.001 & 0.427 &  \\
VC26 & 266.5836597 & -28.8078707 & 19.313 & 16.372 & 14.693 &  &  & 1.123 &  \\
VC27 & 266.5734602 & -28.8016825 & & 16.785 & 14.139 &  &  & 0.534 &  \\
VC28 & 266.5862002 & -28.8080830 & & 16.869 & 14.226 &  &  & 0.642 & 506.9 \\
VC29 & 266.5768727 & -28.8028663 & 16.743 & 14.007 & 12.351 & 11.435 & 11.098 & 0.468 &  \\
VC30 & 266.5498139 & -28.8457266 & & 17.048 & 14.338 &  &  & 0.603 &  \\
VC31 & 266.5786450 & -28.8217240 & & 15.201 & 12.057 & 10.749 & 10.261 & 0.254 & 525.1 \\
VC32 & 266.5719952 & -28.8276493 & 17.771 & 15.185 & 12.837 &  &  & 0.718 &  \\
VC33 & 266.5859465 & -28.8323713 & & 15.850 & 13.411 &  &  & 0.493 &  \\
\hline
\end{tabular}
\end{table*}

\subsection{Color-Magnitude Diagram}
\label{cmd}

\begin{figure*}
\includegraphics[width=\textwidth]{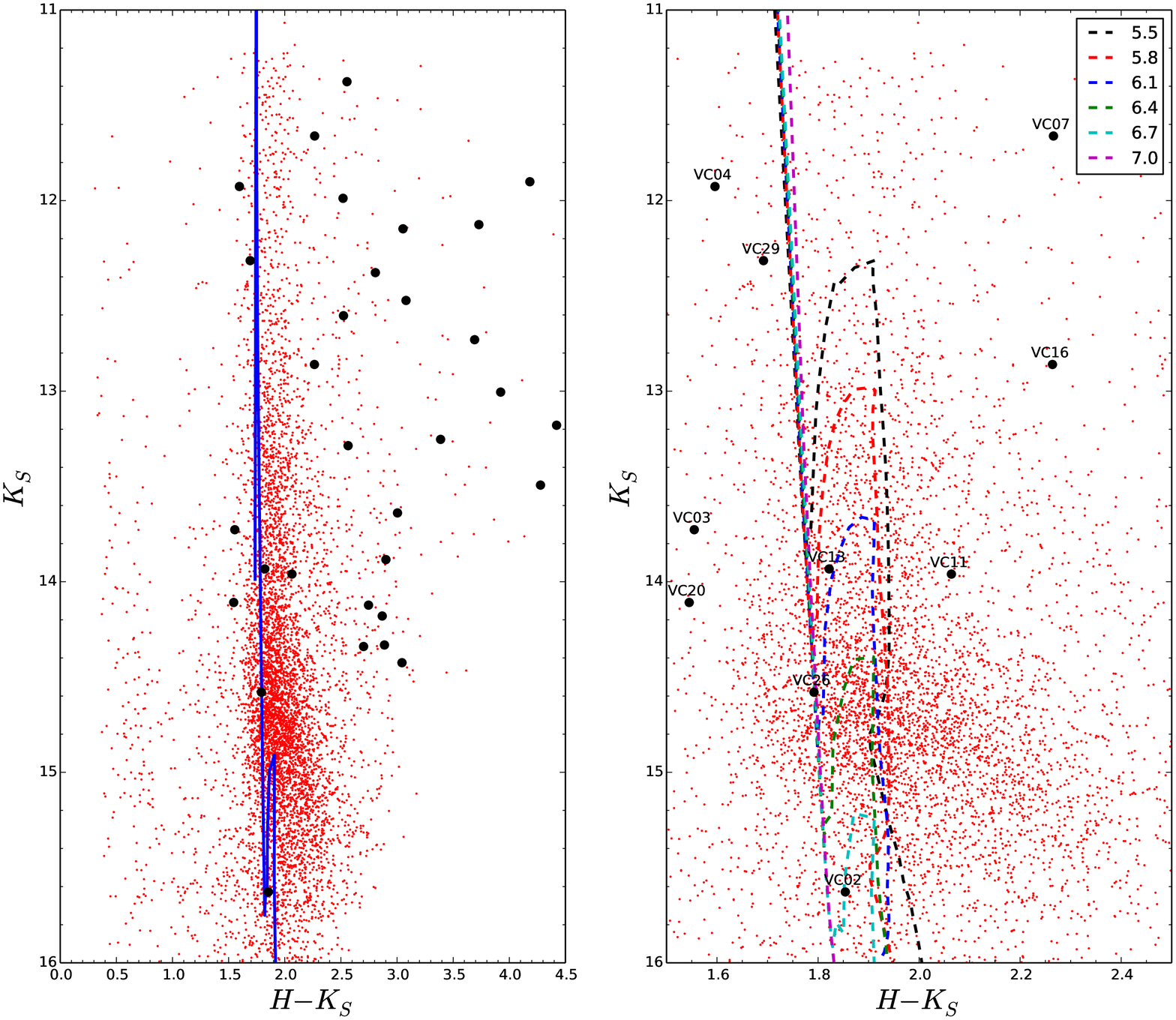}
\caption{{\it Left panel:} Color-magnitude diagram for targets within $2'$ from Quintuplet center. The positions of our variable candidates are marked as black circles. A 4 Myr MS isochrone combined with a PMS isochrone is shown as a blue line. Isochrones were shifted assuming a reddening value of $E(H-K_S)=1.85$. It is important to note that the $K_S$ magnitude used for the CMD corresponds to the first epoch, which is where all other single band measurements were carried out. {\it Right panel:} PMS tracks overplotted for five different ages, from $\log (t/{\rm yr}) = 5.5$ to $7.0$. labels have been added to the variable candidates in the region to help with their identification. Four candidates lie close to the PMS region, VC02, VC11, VC13 and VC26. All of these might be young stellar objects, though further research must be carried out to confirm this possibility. \label{fig:cmd}}
\end{figure*}

Objects found within $2'$ from the cluster center were matched with the single epoch photometries in $J$ and $H$. We noted that only a small fraction of these objects had a $J$-band measurement, due to the strong reddening present in this field. This was already predicted by the results found in Section~\ref{compl}. Hence we decided to utilize the $H$-band to build a reliable color-magnitude diagram (CMD). The resulting diagram is shown in Figure~\ref{fig:cmd}. In addition to the CMD, we included a 4 Myr isochrone, comprised by a PARSEC main sequence (MS) \citep{bre12} and a PISA pre-main sequence (PMS) isochrone, based on the interior model by \citet{tog11}, for $M < 5 M_{\odot}$, both transformed into the VISTA photometric system. The isochrones are shifted assuming a distance of 8 kpc \citep{ghe08}. A fit of the isochrones to the CMD gives us a reddening value of $E(H-K_S)=1.85$.

Since we are including a radius twice the size of the cluster, we expect the CMD to be strongly populated with field stars. Moreover, the central region of Quintuplet is heavily contaminated by bright stars, even in the $K_S$-band. If we add that stars with $K_S<11$ will probably be saturated in the VVV images, we can expect that the fraction of cluster stars that are present in our photometric catalog should not be too large. To prove this, we cross-checked our catalog with the one found in the proper-motion study by \citet{sto15}. Within 1.7 (5$\times$0.34) arcsec, we found only 649 matches, of which 108 are likely members, 167 are likely non-members, while the rest remain uncertain. The small number of counterparts is a consequence of the cluster not being fully resolved in the VVV images, while {\em Hubble Space Telescope} images used by \citet{sto15} allowed them to resolve the cluster down to its core. When we cross-check with our candidate list, we see that the four variables that are within their covered region, namely VC08, VC12, VC14 and VC32, are present. However, VC32 is likely a non member, while the other three have an uncertain membership. With this background we conclude that most of our variable candidates, represented as black circles in the CMD, are not cluster members. See Section~\ref{disc} for a more in-depth discussion of the candidates found.

%------------------------------------------------------------------------------

\section{Discussion}

\subsection{Comparison with Previous Works}

As mentioned in Section~1, only a few variability searches have been carried out in Quintuplet. From our sample, nine candidates had been previously discovered. Hence, 24 new variable candidates have been found by the present work. The results are summarized in Table~\ref{tbl-2}. The first column contains the name used in the present paper. Column 2 includes the number used in the respective catalog: [MKN09] stands for \citet{mat09}, and [MFK13] for \citet{mat13}. Columns 3 and 4 compare periods found by our present work and previous papers, respectively, while column 5 delivers the variability type of the candidate.

\begin{figure*}
\includegraphics[width=\textwidth]{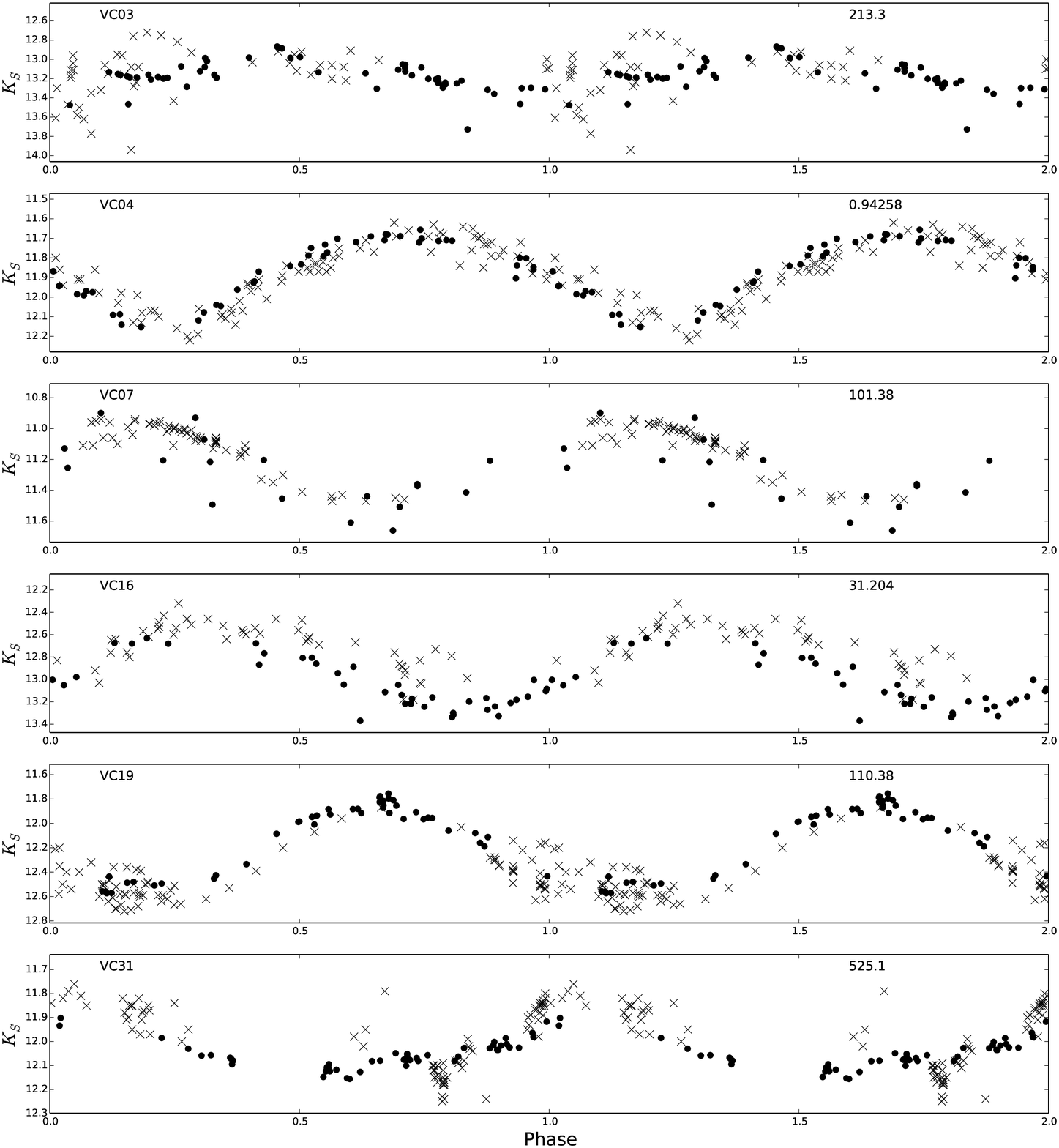}
\caption{Combination of epochs from VVV (circles) and MKN09, MKF13 (crosses) for six variable candidates. Except by VC03 and VC31, periods were fitted to the combined data, and they are similar to those found individually in table 2. For those two particular cases, we kept the original period found by this work, since there is no agreement between both data sets.\label{fig:comp}}
\end{figure*}

In Figure~\ref{fig:comp} we combined our data with the information available from MKN09 and MFK13. The agreement with the two variables from \citet{mat13} is remarkable, with both periods being almost equal. For the Miras of MKN09 we found a good agreement, specially for VC19. Additionally, we managed to estimate periods for VC03 and VC31, while MKN09 could not find any. VC03 is worth discussing, since this is the candidate with the most notorious discrepancy with MKN09. We cannot attribute these differences to photometric errors, thus a possible explanation is that the star has undergone changes in the nature of its variability during the past years. As a consequence, we tentatively classify this variable as a slow irregular or semiregular. The other three candidates (VC23, VC25 and VC32) do not show a clear periodicity, even with the combined data, so we also classify them as slow irregular or semiregular variables.

%\begin{deluxetable}{rcccc}
%\tablecolumns{4}
%\tablewidth{0pt}
\begin{table}
\centering
\caption{Previously discovered variables}
\label{tbl-2}
\begin{tabular}{rcccl}
\hline
ID &  Prev. ID &  P &  Prev. P &  Type\\
\hline

VC03 & [MKN09] 1172 & 213.3 &  & Mira\\
VC04 & [MFK13] 37 & 0.94255 & 0.94255 & Ecl \\
VC07 & [MKN09] 1112 & 103.1 & 102 & Mira \\
VC16 & [MFK13] 39 & 31.29 & 31.279 & Ceph(II) \\
VC19 & [MKN09] 1250 & 110.9 & 108 & Mira \\
VC23 & [MKN09] 1194 & & & Mira \\
VC25 & [MKN09] 1286 & & & Mira \\
VC31 & [MKN09] 1260 & 525.1 & & Mira \\
VC32 & [MKN09] 1236 & & & Mira \\
\hline
\end{tabular}
\end{table}

\subsection{GLIMPSE data}

\begin{figure}
\includegraphics[width=\columnwidth]{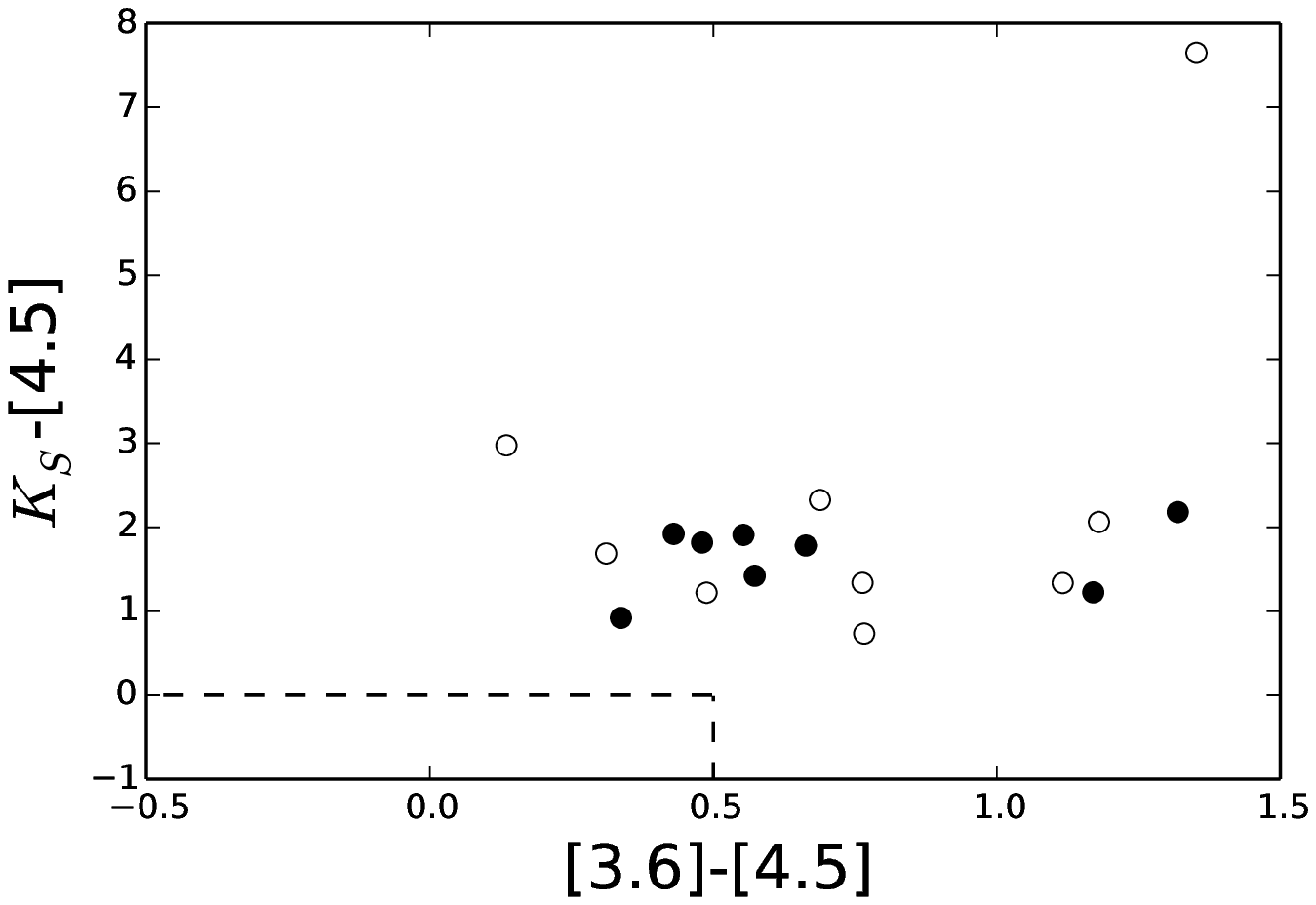}
\caption{Color-color diagram for the variable candidates found within the GLIMPSE database. Periodic variables are plotted as open circles, while black circles correspond to non-periodic stars. The dashed lines enclose the region of MS stars.\label{fig:glimpse}}
\end{figure}

Since one of our long-term goals is to detect young stellar objects, we cross-checked our sample with the magnitudes of GLIMPSE \citep{ben03,chu09}, in order to search for infrared excess. The data was obtained from the GLIMPSE Source Catalog (I + II + 3D), directly from the IRSA website\footnote{http://irsa.ipac.caltech.edu/data/SPITZER/GLIMPSE}. The cross-match found magnitudes for 17 of our objects, with a tolerance of $1''$. The resulting color-color diagram is shown in Figure~\ref{fig:glimpse}. The point in the upper right corner corresponds to VC17, an extremely red star. This is a peculiar object, with a very large $K_S$ amplitude. Unfortunately, it does not appear in our $H$ or $J$ photometry, since at the epoch of these measurements the star was in a faint phase, hence no additional information could be obtained from the CMD. All candidates with a matched GLIMPSE photometry show some degree of infrared excess. Five of these objects have $[3.6]-[4.5]>1$, these are VC06, VC15, VC19 and VC23, besides the already mentioned VC17.

\subsection{Analysis of individual candidates}
\label{disc}

As mentioned in Section~\ref{cmd}, our variable candidates do not seem to belong to the cluster population. In addition, an analysis of the CMD position and shape of their light curves led us to conclude that the majority of these candidates correspond to long-period variables (LPV), including Miras and semiregular variables, which are typically oscillating red giants \citep[see, e.g.,][]{cat15}. If we take into account the young age of the Quintuplet cluster, we do not expect to find a significant number of red giant members. In addition to the Miras already discovered by MKN09, VC06 and VC17 can be classified as Miras too. For VC01 and VC07, periodicity is not well defined, thus they are classified as semiregular variables.

Other variables like VC05, VC08 and VC12 either have very long periods (over $\sim 900$ days) or are semiregular, meaning they are probably OH/IR stars. A larger time coverage would be required to answer this question. In any case, since these are evolved stars, they are most likely not related to the cluster population.

Given its position on the CMD, VC02 seems a very strong YSO candidate. Located in the PMS branch of the 4~Myr isochrone, its light curve shows some sudden brightness increases, the most important being around $\sim 1000$ days. These episodes are consistent with the bursts in the accretion rate found in PMS stars. If we assume it is a cluster member, the position in the CMD suggests a mass of $\sim 2 M_{\odot}$. Since we lack $J$-band and GLIMPSE photometry for this object, no color-cut could be performed to classify it. VC13 and VC26 also qualify as YSO candidates, given their positions in the CMD and the shape of the light curves. A similar situation occurs with VC21. Its light curve show strong, irregular bursts that resemble a YSO, and its $K_S$ magnitude and position in the field makes us think it is a young Quintuplet member. However, no $H$ {\bf or} $J$ information is available for that candidate. For all of these objects, spectroscopic follow-up is required to confirm them as YSOs. The case of VC11 is not as clear as for the other candidates discussed. As observed in the right panel of Figure~\ref{cmd}, the object lies too far from the PMS tracks, even from the youngest one. The light curve is classified as that of an irregular variable star.

VC01, VC09, VC22 and VC28 are classified as semiregular variables, since there is a resemblance of periodicity in all of their light curves. The remainder of the non-periodic variables are difficult to classify. Some of them, for example VC15 and VC24, closely resemble classical T-Tauri light curves \citep[see e. g.][]{mcg15}, but are too bright to be YSOs belonging to the cluster. However, they could be pre-main sequence field stars.

\section{Summary}

In this paper we performed an extensive variability study of the Quintuplet cluster using the five year database of Vista Variables in the Via Lactea ESO Large Survey in the near-infrared.

\begin{itemize}

	\item PSF Photometry with DoPHOT has been performed for $K_S$-band images obtained from the VVV Survey in the region of the Quintuplet cluster. A total of 7586 objects were identified in the zone around 2' from the cluster center. In addition, light curves obtained with DIA were added to our sample.

	\item Two different selection criteria were used to find photometric variation in these objects. After rejecting false positives through visual inspection, and cross-matching objects obtained through DoPHOT and DIA, the final sample was reduced to 31 candidates from DoPHOT and 2 candidates from DIA.
	
	\item A single epoch $H$-band image was used to construct a CMD. An isochrone fitting allowed us to identify three probable pre-main sequence cluster members. 
	
	\item A comparison with previous studies shows that 9 of our candidates were already found in the literature. Hence, in this research we present 24 new variable stars.
	
	\item Through analyzing the light curve shape, and the period obtained in some cases, we conclude that most of our periodic variable stars are LPVs of the Mira type, and some others OH/IR stars with very long period or slow/semiregular variables. Additional epochs from the final VVV Data Release should help to solve this ambiguity.
	
	\item VC02, VC13, VC21 and VC26 have light curves and a position in the CMD (with the exception of VC21) that resemble young stellar objects. The lack of $H$ or $J$-band photometry for some of these objects prevents a further analysis without an additional spectroscopic follow-up.

\end{itemize}

\section*{Acknowledgements}

The authors would like to thank the anonymous referee for the helpful comments and suggestions. Support for CNM, JB, JA-G, RK, MC, NM and ID is provided by the Ministry of Economy, Development and Tourism's Millennium Science Initiative, through grant IC120009, awarded to the Millennium Institute of Astrophysics (MAS). CNM is supported by proyecto GEMINI-CONICYT No. 32110004 and ALMA-CONICYT No. 31120015. We gratefully aknowledge additional support by Proyecto Basal PFB-06/2007 and by proyecto FONDECYT Regular \#1141141 (MC), and FONDECYT Regular \#1130196 (DM). J.A-G. also aknowledges support from FONDECYT Iniciaci\'on 11150916. We finally aknowledge support by CONICYT REDES Project \#140042.

%--------------------------------------------------------------------------------------------------------------------------------
\appendix

\section{Finding Charts}

$K_S$-band finding charts for all candidates were obtained from the VSA\footnote{VISTA Science Archive, http://horus.roe.ac.uk/vsa/}.

\begin{figure*}
\includegraphics[width=0.8\textwidth]{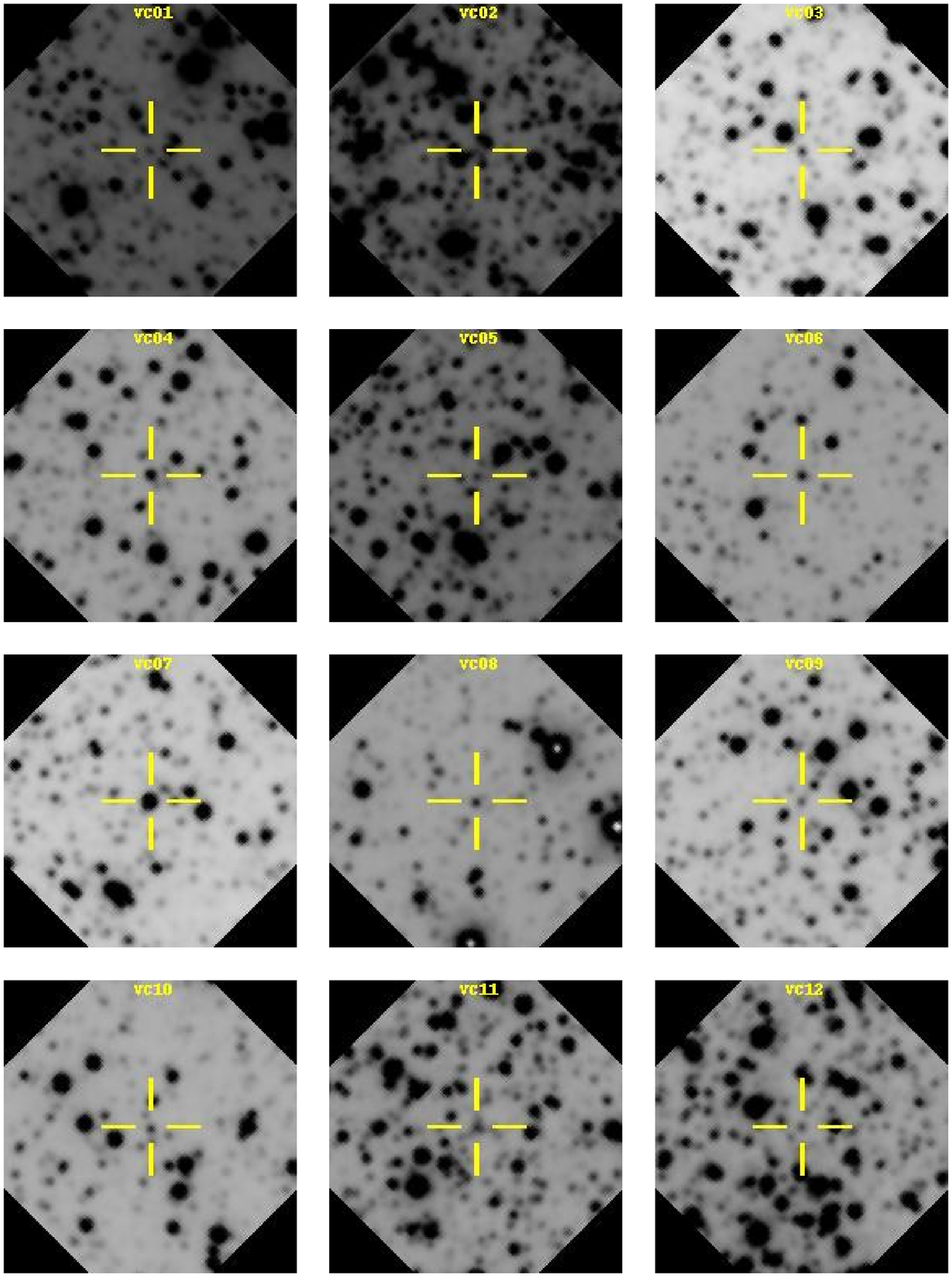}
\caption{$K_S$-band finding charts for all variable candidates. Each thumbnail is $0.5'\times 0.5'$. Orientation is north up, east left, for all charts.}
\end{figure*}

\clearpage

\begin{figure*}
\includegraphics[width=0.8\textwidth]{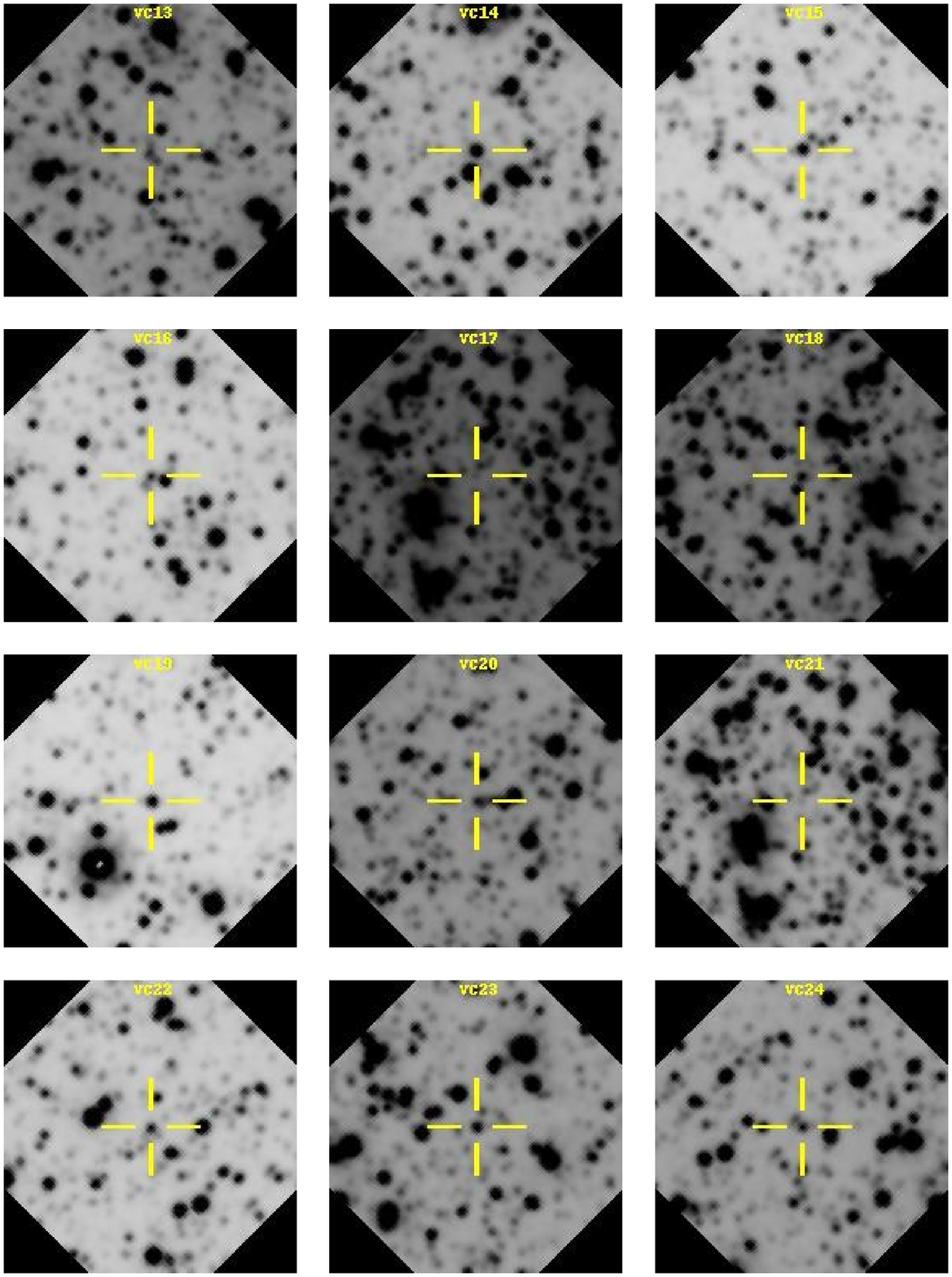}
\contcaption{}
\end{figure*}

\clearpage

\begin{figure*}
\includegraphics[width=0.8\textwidth]{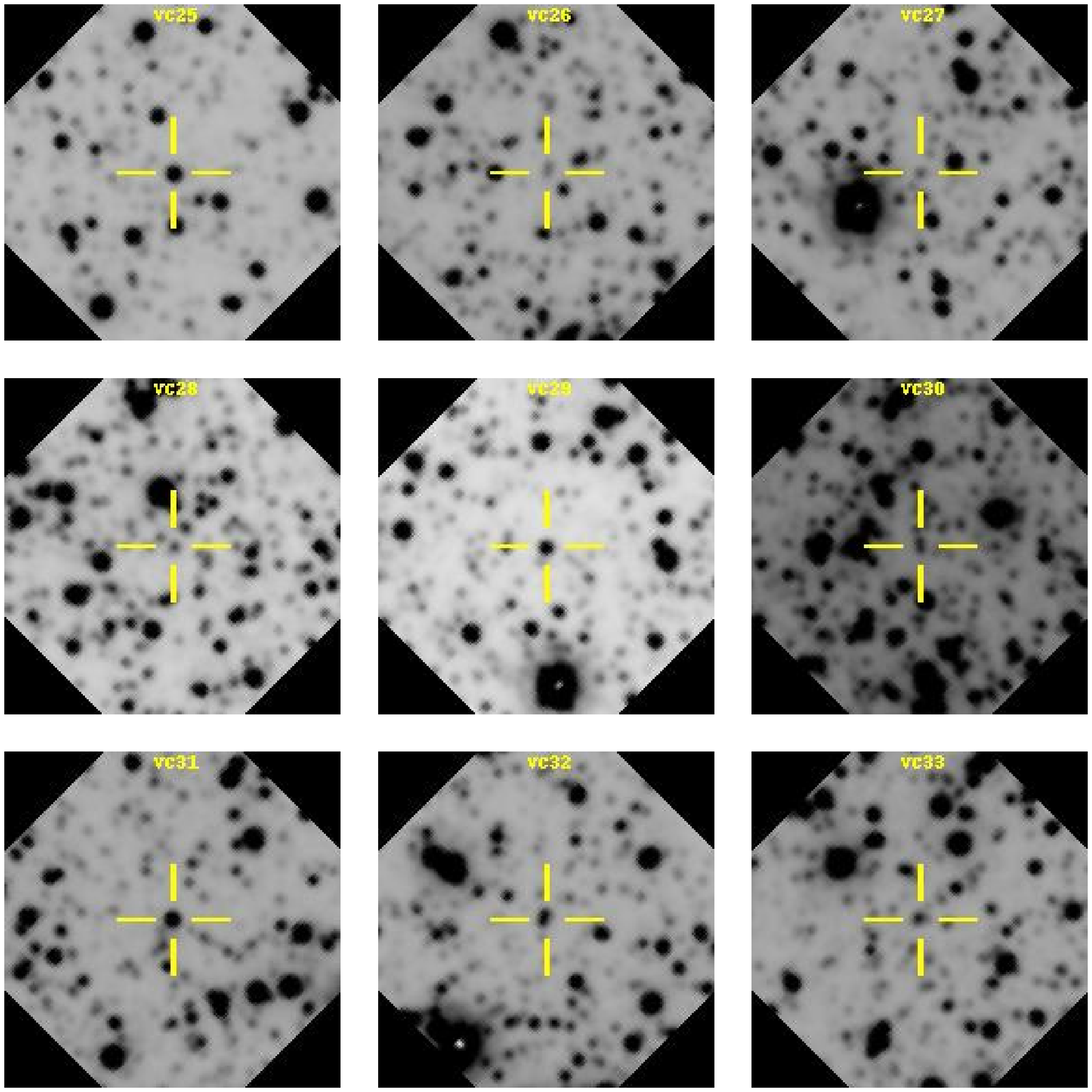}
\contcaption{}
\end{figure*}

\label{lastpage}
\end{document}